# A Collaborative Filtering Based Approach for Recommending Elective Courses

Sanjog Ray, Anuj Sharma

Information Systems Area,
Indian Institute of Management Indore,
Indore, Madhya Pradesh, India
{sanjogr, f09anujs}@iimidr.ac.in

**Abstract.** In management education programmes today, students face a difficult time in choosing electives as the number of electives available are many. As the range and diversity of different elective courses available for selection have increased, course recommendation systems that help students in making choices about courses have become more relevant. In this paper we extend the concept of collaborative filtering approach to develop a course recommendation system. The proposed approach provides student an accurate prediction of the grade they may get if they choose a particular course, which will be helpful when they decide on selecting elective courses, as grade is an important parameter for a student while deciding on an elective course. We experimentally evaluate the collaborative filtering approach on a real life data set and show that the proposed system is effective in terms of accuracy.

**Keywords:** Course Recommender System, Collaborative Filtering, User based Collaborative Filtering, Item based Collaborative Filtering.

## 1 Introduction

The structure, content and delivery of post graduate programmes within management schools in India has been reorganized from time to time and is now delivered on the basis of a fully modular, semester and credit-based curriculum in many of the top business school. While this new semester or trimester (terms as in the IIMs[1]) approach provides increased choices and flexibility and allows students the ability to personalize their studies, challenges have arisen with regard to enabling students to appreciate the range and diversity of modules (courses) in each term or semester that are available to them. In particular, the current enrolment system makes it difficult for students to locate course options that might best fit their individual niche interest. Due to the diversity of different electives available, students find it difficult and time consuming to select the courses they will like and at the same time can get relatively better grades.

---

[1]The Indian Institutes of Management (IIMs), are graduate business schools in India that also conduct research and provide consultancy services in the field of management to various sectors of the Indian economy.

Students pursuing higher education degrees are faced with two challenges: a myriad of courses from which to choose, and a lack of awareness about which courses to follow and in what order. It is according to their friends and colleagues' recommendations that the many of them select their courses and register accordingly. It would be useful to help students in finding courses of interest by the intermediary of a recommender system.

Recommender systems implement advance data analysis techniques to help users find the items of their interest by producing a predicted likeliness score or a list of top-N recommended items for a given active user. Item recommendations can be made using different methods where each method is having different results. Collaborative filtering (CF) based algorithms provides item recommendations or predictions based on the opinions of other like-minded users.

In other domains, the benefits of deploying recommendation systems to assist users in finding relevant items is well understood and researcher are finding different usage of recommender systems in generating recommendations for different category of items [1, 2]. More recently, research has been conducted into developing such technology for course recommender systems.

In this paper, we present our work on developing collaborative filtering based course recommendation system for integration into business school's existing enrolment system either online or offline. We believe collaborative filtering approach provides student an accurate prediction of the grade they may get if they choose a particular subject based on their performance in earlier courses. The prediction of the grade will be very helpful when students decide on selecting elective courses, as grade is an important parameter for a student while deciding on elective courses.

This paper is organized as follows. Section 2 describes related work done in the area of course recommendation and motivation for the proposed work. Section 3 describes different collaborative recommendation algorithms which can be used to facilitate the course recommendation and enrolment process. These algorithms are empirically evaluated in Section 4. The results are discussed in Section 5 and conclusions are presented in Section 6.

## 2   Related Work

From the last decade, Recommendation System (RSs) have been widely developed, implemented and accepted for various categories of application like recommendation of products (e.g., books, music, movies) and of services (e.g., restaurants, hotels, websites), likewise research has been conducted into developing such technology for course recommender systems.

SCR [3], which is an acronym for Student Course Recommender, suggests courses by using a strategy based on Bayesian Network Modeling. The SCR network learns from the information stored about the students who have used the system. It requires the presence of enough cases in the student database. Therefore, if a user has not started or completed any courses, and is not pursuing any degree at the university, SCR cannot give him any course recommendation.

The Course Recommender System [4] is based on the several different collaborative filtering algorithms like user-based [5], item-based [6], OC1 [7], and a modified variant of C4.5 [8]. The system can predict the usefulness of courses to a particular student based on other users' course ratings. To get accurate recommendations, one must evaluate as many courses as possible. Based on the evaluation results, the author suggests C4.5 as the best algorithm for course recommendation. The system cannot predict recommendations for students who have not taken any courses at the University.

PEL-IRT stands for Personalized E-Learning system using item response theory [9]. It recommends appropriate course material to students, taking into account both course material difficulty and student ability. When using PEL-IRT, students can select course categories and units and can use appropriate keywords to search interesting course material. Once the course material has been recommended to students and they have browsed through, the system asks them to answer two questionnaires. This explicit feedback is used by PEL-IRT to re-evaluate the students' abilities and adjust the course material difficulty used in the recommendation.

The AACORN system, which stands for Academic Advisor Course Recommendation Engine, applies a case-based reasoning approach to course recommendation, has been proposed in [10]. The AACORN system recommends courses to graduate students at De-Paul CTI. The system uses the experience of past students and their course histories as the basis for course advising. In order to determine the similarities between course histories, the system uses a metric commonly used in bio-informatics called the edit distance. The system requires a partial history of the courses followed by a student before it can provide useful recommendations.

CourseAgent is a community-based recommendation system that employs a social navigation approach to deliver recommendations for courses based on students' assessment of their particular career goals [11]. The main theme of this approach is to obtain students' explicit feedback implicitly, as part of their natural interaction with the system. The basic and obvious benefit of the system to the students is as a course management system that keeps information about courses they have taken and facilitates communication with their advisors.

The RARE, a course recommender system based on association rules combines association rules together with user preference data to recommend relevant courses [12]. RARE was used on real data coming from the department of Computer Science at the Universit´e de Montr´eal. It analyses the past behavior of students concerning their course choices. More explicitly, it formalizes association rules that were implicit before. These rules enable the system to predict recommendations for new students. A solution to the cold start problem, which is a central question in recommender systems, is also proposed in RARE.

The Course Recommender System [13] is based on variation on the widely-used item-based collaborative filtering algorithm. The objective of module recommender system is to facilitate and enhance the on-line module selection process by recommending elective modules to students based on the core modules that they have selected. Evaluation using historical enrolment data shows very encouraging performance in terms of both recall and coverage.

Some recent research is focused on using course recommender systems in niche area like for civil engineering professional courses [14] and for university physical education [15].

From the review of the literature, it is evident that recommendation technology applied in education field can facilitate the teaching and learning processes. Considering the significance and seriousness of education, the help of recommendation system can improve efficiency and increase veracity of learners in the actual situation.

Comparing to other approaches like SCR [3] based on bayesian network modeling, RARE based on association rules, and AACORN [10] based on case-based reasoning, the proposed approach uses collaborative filtering as in Course Recommender System [4] but using students' grades that is indicator of performance in earlier courses. The other systems like PEL-IRT [9] and CourseAgent [11] are explicit feedback based system but the proposed approach in this paper does not need any feedback from students. Given the challenges and constraints of integrating this technology into an existing live environment, the proposed work is in its initial stages but the vast literature suggests that this domain offers great potential and scope for future research and development.

## 3   Collaborative Filtering Methods

Collaborative filtering methods are those methods that generate personalized recommendations based on user preferences data or judgment data on different items present in the system. Judgment data is primarily in form of ratings assigned by users to different items. Ratings data can be explicitly or implicitly obtained. Explicit ratings are those given directly by the user, while implicit data can be collected indirectly by studying data about the user from different sources like purchase data, browsing behavior etc.

Collaborative filtering (CF) was first introduced by to recommend jokes to users [16]. Since then many systems have used collaborative filtering to automate predictions and its applications in commercial recommender systems has resulted in much success [17]. Because of its minimal information requirements and high accuracy in recommendations, variations of CF based recommender systems have been successfully implemented in Amazon.com [1], TiVo [18], Cdnow.com [19], Netflix.com [20] etc. The most widely used approach in collaborative filtering are the nearest neighbors approach. We describe below two of the most popular collaborative filtering approaches, user-based collaborative filtering and item-based collaborative filtering.

### 3.1   User-Based Collaborative Filtering

User-based collaborative filtering was first introduced by GroupLens research systems [21, 22] to provide personalized predictions for Usenet news articles. The basis implementation details of user–based CF remains the same as proposed in [22]. CF systems are primarily used to solve the prediction problem or the top-N prediction problem. For an active user $U_a$ in the set of users $U$, the prediction problem is to predict the rating active user will give to an item $I_t$ from the set of all items that $U_a$ has not yet rated. The steps followed in user-based CF to make a prediction for user $U_a$ are as follows:

Step 1: Similarity between the active user $U_a$ and every other user is calculated.

Step 2: Based on their similarity value with user $U_a$, set of $k$ users, most similar to active user $U_a$ is then selected.

Step 3: Finally, prediction for item $I_t$ is generated by taking the weighted average of the ratings given by the $k$ similar neighbors to item $I_t$.

In step 1 to calculate the similarity between users Pearson-r correlation coefficient is used. Let the set of items rated by both users $u$ and $v$ be denoted by $I$, then similarity coefficient ($Sim_{u,v}$) between them is calculated as

$$Sim_{u,v} = \frac{\sum_{i \in I}(r_{u,i} - \bar{r_u})(r_{v,i} - \bar{r_v})}{\sqrt{\sum_{i \in I}(r_{u,i} - \bar{r_u})^2} \sqrt{\sum_{i \in I}(r_{v,i} - \bar{r_v})^2}} \quad (1)$$

Here $r_{u,i}$ denotes the rating of user $u$ for item $i$, and $\bar{r_u}$ is the average rating given by user $u$ calculated over all items rated by $u$. Similarly, $\bar{r_v}$ denotes the rating of user $v$ for item $i$, and $\bar{r_v}$ is the average rating given by user $v$ calculated over all items rated by $v$. In some cases to calculate similarity cosine vector similarity is used. In [5] through experimental evaluation they have shown the using person correlations results in better accuracy. To improve accuracy of predictions, various methods have been proposed to improve the way similarity is measured between users. Significance weighting [23] and case amplification [5] are two methods that have been experimentally shown to impact accuracy positively. Case amplification transforms the correlation value when used as weight-age in step 3. Correlations closer to one are emphasized and those less than zero are devalued. Significance weighing is applied to devalue correlation value calculated between two users based on a small number of co-rated items. When the numbers of co-rated item are above a particular threshold it has no impact on actual correlations value. In [23] the threshold has been shown to be 50, but threshold value may differ among datasets, so it should be experimentally determined.

Once similarities are calculated, a set of users most similar to the active user $U_a$ are selected in step 2. There are two ways in which a set of similar users can be selected. One is to select all users whose correlation with user $U_a$ lie above a certain specified correlation value or select a set of top-$k$ users, similarity wise. Experimentally it has been shown that top-$k$ approach performs better than the threshold approach [23]. Value of $k$ is obtained by conducting experiments on the data as it depends on the data set used.

In step 3 to compute the prediction for an item $i$ for target user $u$, an adjusted weight-age sum formula is used to take into account the fact that different users have different rating distributions.

$$P_{u,i} = \bar{r_u} + \frac{\sum_{v \in V} Sim_{u,v}(r_{v,i} - \bar{r_v})}{\sum_{v \in V} |Sim_{u,v}|} \quad (2)$$

Where, *v* represents the set of *k* similar users. While calculating prediction, only those users in set *v,* who have rated item *I*, are considered.

### 3.2 Item-Based Collaborative Filtering

The main difference between item-based CF [6 and 24] and user-based CF is that item-based CF generates predictions based on a model of item-item similarity rather than user-user similarity. In item-based collaborative filtering, first, similarities between the various items are computed. Then from the set of items rated by the target user, *k* items most similar to the target item are selected. For computing the prediction for the target item, weighted average is taken of the target user's ratings on the *k* similar items earlier selected. Weight-age used is the similarity coefficient value between the target item and the *k* similar items rated by the target user. To compute item-item similarity adjusted cosine similarity is used.

Let the set of users who rated both items *i* and *j* be denoted by *U*, then similarity coefficient ($Sim_{i,j}$) between them is calculated as

$$Sim_{i,j} = \frac{\sum_{u \in U}(r_{u,i} - \overline{r_u})(r_{u,j} - \overline{r_u})}{\sqrt{\sum_{u \in U}(r_{u,i} - \overline{r_u})^2} \sqrt{\sum_{u \in U}(r_{u,j} - \overline{r_u})^2}} \quad (3)$$

Here $r_{u,i}$ denote the rating of user *u* for item *i*, and $\overline{r_u}$ is the average rating given by user u calculated over all items rated by *u*. Similarly, $r_{u,j}$ denotes the rating of user *u* for item *j*.

To compute the predicted rating for a target item *i* for target user *u*, we use the following formula.

$$P_{u,i} = \frac{\sum_{j \in I} Sim_{i,j} * r_{u,j}}{\sum_{j \in I} |Sim_{i,j}|} \quad (4)$$

In equation 4, *I* represent the set of *k* most similar items to target item *i* that have already been rated by target user *u*. As earlier mentioned, $r_{u,j}$ denotes the rating of user *u* for item *j*.

## 4   Experimental Evaluation

We performed the experimental evaluation on the anonymized data set of 255 students and the grades they scored in 25 subjects. Among the 25 subjects, semester 1 and semester 2 comprised of 9 subjects each, while the third semester comprised of 7 subjects. Out of the total 6375 data rows, training set comprised of 6200 data rows and test set comprised of 175 data rows. To create the test data set we randomly selected 25 students and separated their grade data for the third semester from the dataset. We evaluated the algorithms for different ratios of test /train data i.e., *x* values. Both item-based and user-algorithms were implemented as described in the earlier section. Figure 1 shows the proposed approach for recommending courses.

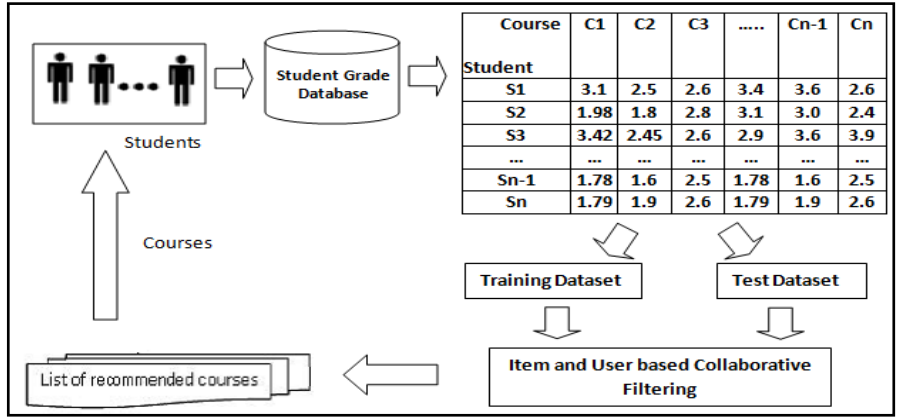

**Fig. 1.** The Proposed Course Recommendation Approach

To measure the recommendations quality we use most prominent and widely used metric for predictive accuracy mean absolute error (MAE) [17], [23], [25]. Mean absolute error (often referred to as MAE) measures the average absolute deviation between a predicted rating and the user's true rating. The MAE is computed by first summing the absolute errors of the N corresponding ratings-prediction pairs and then computing the average. Lower MAE values indicate higher accuracy. MAE is the most widely used metric because the mechanics of its computation are simple and easy to understand. In our experiments we measured the MAE for user-based and item-based algorithms for neighborhood size 5, 10, 15, 20 and all users respectively. In our experiment only those users with positive correlation values with the test user were considered for selecting K-nearest neighbors.

## 5 Experimental Results

In this section, we present the results of our experimental evaluation. Figure 2 and 3 shows the MAE values for user-based CF and item-based CF for different values of $k$ i.e. neighborhood size and different values of $x$ i.e. test/train data ratio. Overall prediction accuracy of both the algorithms is very high as MAE values for all possible neighborhood size falls in the range of 0.33 to 0.38. As we can observe from the results, there is not much difference in accuracy between item-based CF and user based CF algorithms. In item based CF we also observe that for larger values of $k$, MAE values hardly change. The reason for this may be the small number items present in the dataset. Both the algorithms perform worse at $k$=5 and MAE values don't vary much after $k$=10. In figure 4 we compare user-based CF and item based CF different values of $x$ for neighborhood size $(k)$ = 10. While item-based CF perform better for $x$=10%, user-based CF performs slightly better from higher values of $x$.

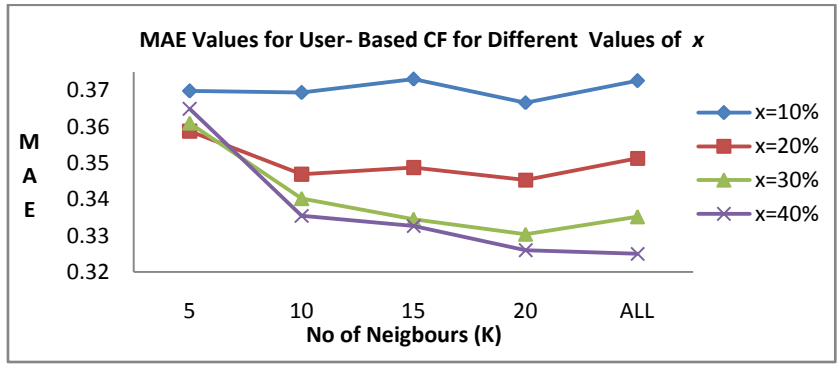

**Fig. 2.** MAE Values for User- Based CF for Different Values of *x*

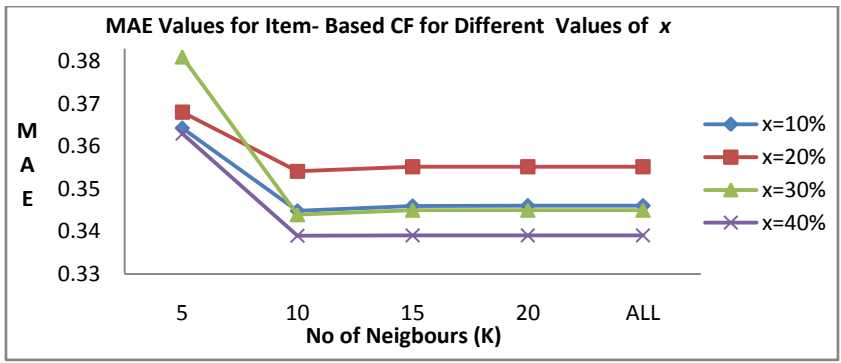

**Fig. 3.** MAE Values for Item- Based CF for Different Values of *x*

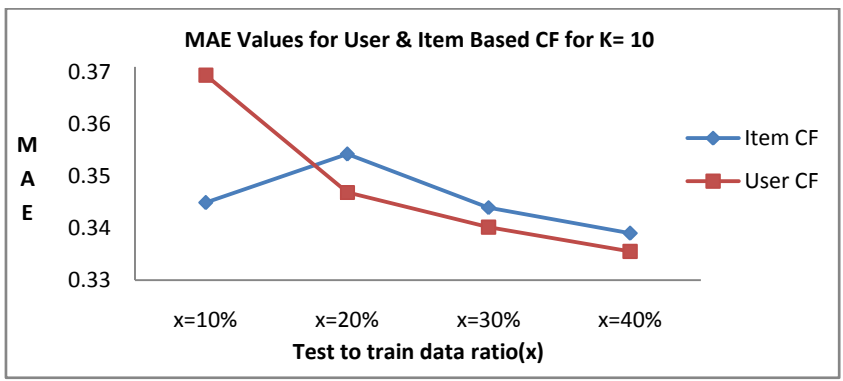

**Fig. 4.** MAE Values for User & Item CF for Neighborhood Size *k* =10

## 6   Conclusion

In this paper, we have compared two collaborative filtering approaches for predicting the grade a student will obtain in different courses based on their performance in earlier courses. Overall, the results of our experimentation on a real life dataset are very encouraging. We believe collaborative filtering approach provides student an accurate prediction of the grade they may get if they choose a particular subject, which will be very helpful when they decide on selecting elective courses, as grade is an important parameter for a student while deciding on elective courses.

For future work, research can done in developing integration strategies for approaches that can accurately predict student performance in courses and approaches that help a select a subject or courses based on student interests and learning objectives. These approaches can be used to provide valuable advice to students during career guidance advice and courses selection process.